\documentclass[aps,prl,twocolumn,groupedaddress]{revtex4-1}
\usepackage{epsfig,graphics}
\usepackage{graphicx}
\usepackage{ifpdf}
    
\begin{document}

\title{Ultra-sensitive detection of mode splitting in active optical microcavities}

\author{Lina He}
\email{heln@ese.wustl.edu}
\affiliation{Department of Electrical and Systems Engineering, Washington University, St. Louis, Missouri 63130, USA}
\author{\c{S}ahin Kaya \"{O}zdemir}
\email{ozdemir@ese.wustl.edu}
\affiliation{Department of Electrical and Systems Engineering, Washington University, St. Louis, Missouri 63130, USA}
\author{Jiangang Zhu}
\affiliation{Department of Electrical and Systems Engineering, Washington University, St. Louis, Missouri 63130, USA}
\author{Lan Yang}
\affiliation{Department of Electrical and Systems Engineering, Washington University, St. Louis, Missouri 63130, USA}

\date{\today}

\begin{abstract}
Scattering induced mode splitting in active microcavities is demonstrated. Below the lasing threshold, quality factor enhancement by optical gain allows resolving, in the wavelength-scanning transmission spectrum, the resonance dips of the split modes which otherwise would not be detected in a passive resonator. In the lasing regime, mode splitting manifests itself as two lasing modes with extremely narrow linewidths. Mixing of these laser modes in a detector leads to a heterodyne beat signal whose frequency corresponds to the amount of splitting. Lasing regime not only allows ultrahigh sensitivity for mode-splitting measurements but also provides an easily accessible scheme by eliminating the need for wavelength scanning around resonant modes. Mode splitting in active microcavities has immediate impact in enhancing the sensitivity of sub-wavelength scatterer detection and in studying light-matter interactions in strong coupling regime.
\end{abstract}


\maketitle

Ultra-high quality factor $\emph{Q}$ and tight confinement of photons in an extremely small mode volume cause strong interactions between cavity photons and matter, making whispering gallery mode (WGM) optical microcavities a suitable platform for sensing applications and for studying fundamental physical problems \cite{Vahala_cavities}. Such cavities are, in particular, very attractive for cavity quantum electrodynamics (cavity QED) studies. In weak coupling regime, the system is dominated by an irreversible decay of the dipole, whereas in strong coupling regime the system undergoes a reversible dynamics which allows coherent exchange of a quantum between the dipole and the cavity mode leading to Rabi oscillations. When the coupling strength exceeds the mean of the decay rates, mode splitting is observed. This regime has been demonstrated using trapped atoms and quantum dots interacting with WGMs \cite{VCR1,Vahala192345}.

The modal coupling in an ultra-high-$\emph{Q}$ WGM microcavity induced by a Rayleigh scatterer provides an analogy to cavity QED effects. Single Rayleigh scatterer mediated coupling between the degenerate clockwise (CW) and counter-clockwise (CCW) propagating WGMs lifts their degeneracy resulting in mode splitting which can be observed as a doublet in the transmission spectrum. Modal coupling was first reported in ultra-high-$\emph{Q}$ microspheres by Il'chenko and Gorodetsky \cite{Gorodetskii} and later addressed in detail by Weiss {\it et al.} \cite{Weiss} leading to intense theoretical and experimental investigations of intrinsic (i.e., due to material inhomogeneities, structural defects, contaminations) and intentionally induced (i.e., due to controlled placement of dipoles in the mode volume) mode splitting in various passive WGM resonators including microspheres, microtoroids and microdisks \cite{Mazzei,Kippenberg,VCR1}. Recently, we showed that intentionally induced mode splitting in a silica microtoroid can be used for detecting, counting and sizing individual nanoparticles deposited onto the resonator \cite{Jiangang}.

Mode splitting can be resolved if the strong coupling condition $2|g|>\Gamma_R+(\omega_0/\emph{Q})$ is satisfied \cite{Jiangang}, i.e., coupling strength exceeds the mean of intrinsic and dipole induced decay rates of the system. Here, $2g=-\alpha f^2 (r)\omega_0/V$ and $2\Gamma_R=-2g\alpha\omega_0^3/3\pi c^3$ represent the differences in resonance frequencies and linewidths of the two split modes, $\alpha$ denotes the polarizability of a dipole located at position $r$ in the mode volume $V$ with normalized field distribution $f(r)$, $\omega_0$ is the resonance frequency of the WGM and $c$ is the speed of light. The strong coupling condition determines the lower bound of detectable mode splitting for a given microcavity. For example, in the limit of $\Gamma_R=0$, angular frequency splitting should be larger than $1.22 ~\rm MHz$ to be resolved at $1550~ {\rm nm}$ with a microcavity of $\emph{Q}\sim 10^9$. Absence of mode splitting in the transmission spectrum implies that either the system is free of scattering centers and hence free of mode splitting or it is not in the strong coupling regime, i.e., $2|g|$ is so small that the splitting cannot be resolved.

Resolving mode splitting requires both a dipole of large $\alpha$ with small $\Gamma_R$ and a cavity of small $V$ with high $\emph{Q}$. For nanoparticle detection, one does not have control on the dipole properties but can fabricate microcavities with high $\emph{Q}/V$. Mode volume $V$ cannot be arbitrarily small because of two reasons: First, decreasing $V$ increases $\Gamma_R$; Second, radiation loss reduces $\emph{Q}$ significantly when $V$ is below some threshold value. However, $\emph{Q}$ can be maximized by minimizing optical losses in the resonator. Here, we show that a portion of the losses can be compensated by incorporating a gain medium inside the cavity which leads to improved resolvability of mode splitting.

In this Letter, we demonstrate, for the first time, scattering induced mode splitting in an active microcavity and introduce a heterodyne detection scheme for its monitoring. Active microcavity allows resolving small mode splitting, which would otherwise go undetected in a passive resonator. Our scheme avoids the use of transmission spectrum via wavelength scan, thus making mode splitting phenomenon more accessible for fabricating on-chip sensors and for studying fundamental physical phenomena in microcavities.

Active microcavities formed by integrating gain medium (e.g., rare-earth ions such as Er, Yb, Nd) into the cavities have been used for on-chip ultralow-threshold lasers \cite{Lan}. Most active microresonators used in our experiments have intrinsic mode splitting which cannot be resolved without pumping the cavity. If the active microcavity is pumped below the lasing threshold, the gain medium compensates for part of the cavity losses depending on the pumping level and gives rise to enhancement of the cavity-$\emph{Q}$ making mode splitting resolvable in the transmission spectrum. On the other hand, if the pump power exceeds the lasing threshold, the generated laser light splits into two modes with resonance frequencies located at $\omega_1$ and $\omega_2$. The superposition of these two laser modes, when detected by a photodetector of sufficient bandwidth, results in a heterodyne beat note $i(t)=P_1+P_2+2\sqrt{P_1P_2}\cos(\Delta\omega t+\Delta\phi)$, where $P_1$ and $P_2$ are the detected powers of the two split modes, $\Delta\omega=|\omega_1-\omega_2|$ and $\Delta\phi$ are the frequency and phase differences between the split modes, respectively. Therefore, the photodetector output varies sinusoidally with the beat frequency $\Delta\omega$. The amount of splitting $\Delta\omega$ can be detected via measuring the beat frequency either directly by a spectrum analyzer or by performing Fourier transform of the beat note signal in time domain. Thus, wavelength scan of the input light is not required.

\begin{figure}\epsfxsize=7.0cm \epsfbox{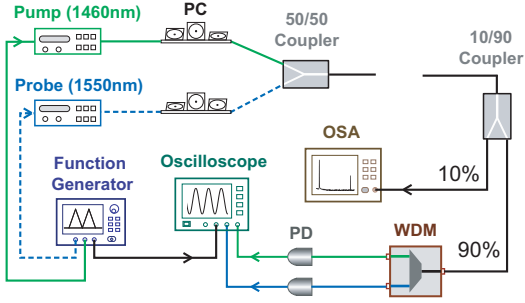} \caption{Experimental setup with a tapered fiber interfaced to an active microtoroid. The dotted lines show the components for the pump-probe experiments. PD: Photodetector, PC: Polarization controller, OSA: Optical spectrum analyzer, WDM: Wavelength division multiplexer.}\label{fig1}\end{figure}

Active Microresonators used in the experiments are Erbium (Er)-doped toroidal microcavities fabricated from Er-doped sol-gel silica films on a silicon wafer \cite{Lan} with diameters $\sim 50~\mu{\rm m}$ and Er concentration of $3\times10^{18}~{\rm ions/}{\rm cm}^{3}$. Schematics of the experimental setup is shown in Fig. \ref{fig1}. Two tunable external-cavity lasers in $1460 ~{\rm nm}$ (pump laser) and $1550 ~{\rm nm}$ (probe laser) bands are used. Spectral band of the probe laser coincides with the emission spectrum of Er and it is used only for pump-probe experiments. The pump wavelength is tuned on resonance with the cavity mode while the probe wavelength is scanned to obtain transmission spectra around the lasing wavelength. In this study, the transmission spectra are taken during up-scan of the probe wavelength. Light is coupled in and out of the microtoroid via a tapered fiber, and the coupling condition is varied by controlling the taper-cavity gap using a nano-positioning system. Light coupled out of the microcavity is divided by a 10/90 coupler, with the $10\%$ port connected to an optical spectrum analyzer (OSA: resolution $0.1 ~{\rm nm}$) and the $90\%$ port directed to a $1460/1550$ wavelength division multiplexer (WDM). After the WDM, light fields are detected by low noise photodetectors (PD; bandwidth: $125~{\rm MHz}$) whose outputs are monitored with an oscilloscope.

\begin{figure}\epsfxsize=8cm \epsfbox{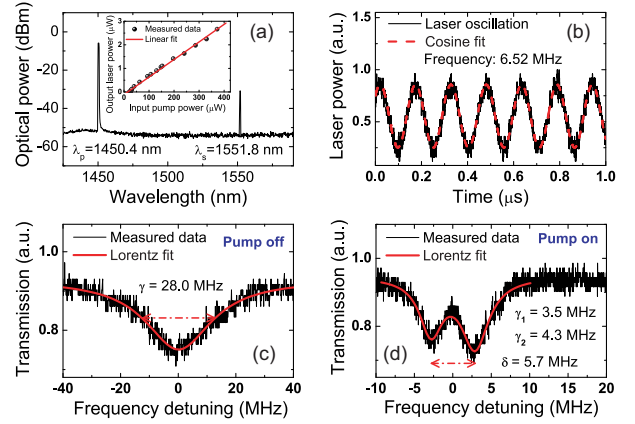} \caption{(a) Lasing characteristics of the active microcavity. (b) Heterodyne beat note signal due to mode splitting in the active microcavity. Transmission spectra are obtained while the probe wavelength is scanned and the pump laser is turned (c) off, and (d) on. $\gamma$: resonance linewidth, $\delta$: amount of doublet splitting. Detuning is the difference between the frequencies of the probe and the cavity resonance.}\label{fig2}\end{figure}

Figure \ref{fig2} shows the lasing characteristics of an Er-doped toroidal microlaser and the transmission spectra of the WGM at the lasing frequency before and after the microcavity is pumped. When the pump power ($\lambda_{\rm p} = 1450.4 ~{\rm nm}$) is above the threshold $P_{\rm th} = 15~\mu{\rm W}$, single mode lasing is observed at $\lambda_{\rm s} = 1551.8 ~{\rm nm}$ with slope efficiency $0.74\%$ by the OSA (Fig. \ref{fig2}(a)). The signal from the PD connected to the $1550 ~{\rm nm}$-band output of the WDM shows a sinusoidal oscillation with frequency $6.25 ~{\rm MHz}$ (Fig. \ref{fig2}(b)) whereas no such oscillation is observed from the other PD connected to the $1460 ~{\rm nm}$-band output of the WDM. This implies that mixing process takes place in $1550 ~{\rm nm}$ band leading to a beat note signal. We confirm this with pump-probe experiments: First, we finely scan the wavelength of the probe laser around $1551.8~ {\rm nm}$ with a scan speed of $42.4 ~{\rm nm/s}$ and record transmission spectra while the pump laser is off. As seen in Fig. \ref{fig2}(c), a single resonance dip (probe mode) with linewidth $28~ {\rm MHz}$ corresponding to $\emph{Q}\sim 6.9\times 10^6$ and photon lifetime $5.7~{\rm ns}$ is observed. Then the experiment is repeated when the pump laser is turned on and a small pump power is coupled into the microcavity. In this case, we observe a doublet in the transmission spectrum for the probe laser (Fig. \ref{fig2}(d)). The two resonances are separated by a frequency of $5.7~{\rm MHz}$, and their linewidths are $3.5~{\rm MHz}$ and $4.3~{\rm MHz}$ corresponding to $\emph{Q}\sim 5.5\times 10^7$ and $\emph{Q}\sim 4.5\times 10^7$, respectively. The observation can be explained as follows. When the pump laser is off, $\emph{Q}$ factor of the probe mode is so small that mode splitting cannot be resolved. This is justified as the estimated splitting $6.52~{\rm MHz}$ is much smaller than the linewidth $28~{\rm MHz}$ of the probe mode when there is no pump. After the pump laser is turned on, part of the cavity losses is compensated by the optical gain leading to a higher $\emph{Q}$ and a narrower linewidth which, in return, reveals the doublet in the transmission spectrum. The discrepancy between the amount of splitting estimated from the transmission spectrum and the beat frequency is attributed to the effect of thermal heating of the resonator during wavelength scan \cite{Tal,Lina}.

\begin{figure}\epsfxsize=8cm \epsfbox{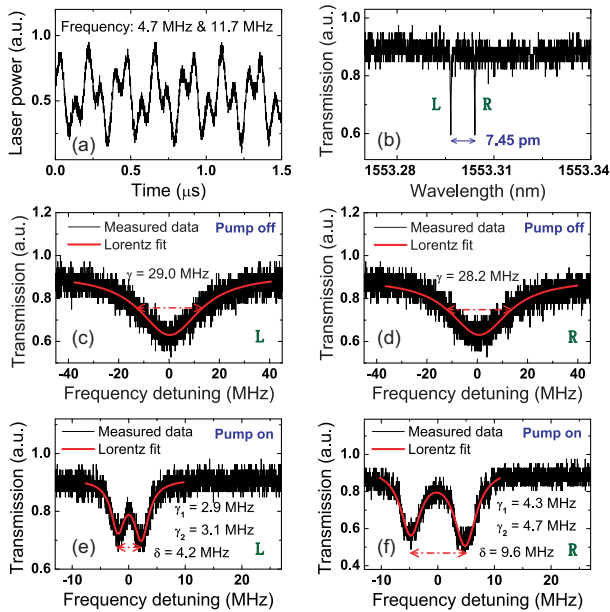} \caption{(a) Heterodyne beat note signal due to mode splitting in a two-mode microlaser. (b) Transmission spectrum when the probe wavelength is scanned with pump laser off. Zoom-in of the resonant modes (c) L and (d) R. Transmission spectra obtained when the probe wavelength is scanned with pump laser on for modes (e) L and (f) R. }\label{fig3}\end{figure}

In an active microcavity with multi-mode lasing, each lasing mode undergoes distinct mode splitting determined by its overlap with the scattering centers (i.e., coupling strength between the WGM field and the scattering centers), and produces its own heterodyne beat signal when detected by a photodetector. We choose an Er-doped microtoroid lasing at two different wavelengths to study the effect of multi-mode lasing on the beat note signal. The detected beat note for this two-mode microlaser is no longer a single frequency sinusoidal oscillation (Fig. \ref{fig3}(a)). Fourier transform of the beat signal reveals two frequency components, $4.7~{\rm MHz}$ and $11.7 ~{\rm MHz}$, suggesting that the two lasing modes experience different mode splittings. We perform pump-probe experiments to further confirm this conclusion. Without pumping the microcavity, transmission spectrum around lasing wavelength shows two adjacent WGMs (Fig. \ref{fig3}(b)) separated by $7.45 ~{\rm pm}$ ($926 ~{\rm MHz}$) with linewidths of $29.0 ~{\rm MHz}$ (Fig. \ref{fig3}(c)) and $28.2 ~{\rm MHz}$ (Fig. \ref{fig3}(d)) with no observable splitting. When the pump laser is turned on, the gain medium starts compensating for the cavity losses and doublet splitting shows up for both WGMs (Figs. \ref{fig3}(e) and (f)). The estimated amount of splitting from the transmission spectra are consistent with the beat frequencies shown in Fig. \ref{fig3}(a). The results indicate that, in case of multi-mode laser, each individual lasing mode experiences different mode splittings depending on its mode distribution. The resultant beat note is a superposition of multiple single-frequency sinusoidal signals and carries information of mode splitting for each lasing mode. The multiple laser mode splitting in an active microcavity can be utilized to decrease measurement errors in detecting and characterizing local changes on and around the resonator (e.g., detecting nanoparticles or dipoles).

\begin{figure}\epsfxsize=8cm \epsfbox{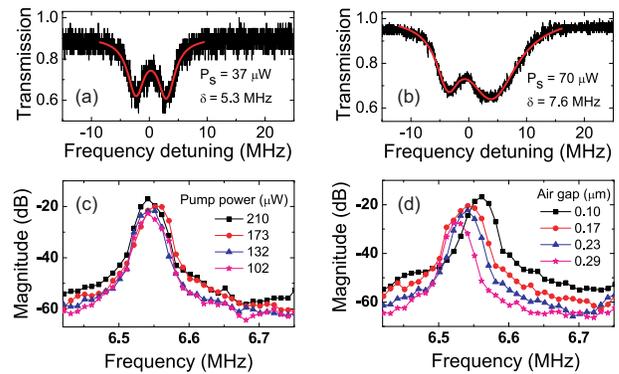} \caption{Transmission spectra of the probe mode for probe power at (a) $37~\mu{\rm W}$ and (b) $70~\mu{\rm W}$. Effects of (c) pump power and (d) taper-cavity gap on Fourier transform of the beat note signal.}\label{fig4}\end{figure}

For the beat note signal to serve as an indicator for mode splitting in active microcavities, it is crucial that the signal is stable against small perturbations. We investigated the effects of optical power and cavity loading condition on the estimated mode splitting in both the wavelength-scan transmission spectrum and the heterodyne measurement schemes. As seen in Figs. \ref{fig4}(a) and (b), increasing the probe power results in a significant change in the mode splitting spectra obtained by scanning the probe wavelength due to thermal heating effect of the resonator. For the heterodyne scheme, on the other hand, as presented in Fig. \ref{fig4}(c) the beat frequency shows no obvious change as the pump power increases, confirming that mode splitting is independent of laser power \cite{Weiss}. Figure \ref{fig4}(d) depicts that the beat frequency decreases slightly with increasing taper-cavity gap, e.g., an increase of $0.2~ \mu{\rm m}$ leads to a change less than $1\%$. These results suggest that heterodyne technique provides more stable readings of mode splitting.

\begin{figure}[]\epsfxsize=8cm \epsfbox{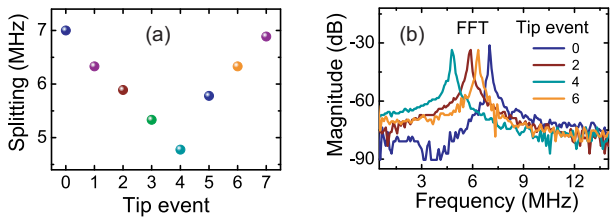} \caption{(a) Effect of an intentionally introduced scatterer (fiber tip) on the mode splitting in an active microcavity. (b) Fast Fourier transform (FFT) of beat signal for different tip positions (scatterer size). Increasing tip event number corresponds to increasing size of the scatterer.}\label{fig5}\end{figure}

Figure \ref{fig5} shows the effect of the size of an external scatterer on mode splitting. In these experiments, a fiber tip (size $< 1\mu{\rm m}$) fabricated by heating and pulling an optical fiber followed by buffered-HF etching is used as an external scatterer. After fixing the lateral position of the fiber tip, we gradually move it close to the peripheral of the microcavity and record the beat note signal for different tip positions. Due to the cone-like shape of the tip, this simulates the increase in the size of the scatterer entering the microcavity mode volume. When the fiber tip is placed far away from the microcavity, the intrinsic mode splitting of $7.0 ~{\rm MHz}$ is observed. As the fiber tip approaches the microcavity, mode splitting (beat frequency) first gradually decreases and then starts increasing after a critical size. This is similar to the observation in Ref. \cite{Jiangang2} and can be explained by the size and position of the tip with respect to the WGM mode volume. For small scatterer size, the resonant mode of the doublet with higher frequency is strongly disturbed and shifted to lower frequency side decreasing the amount of mode splitting, i.e., we observed a decrease of $2.2 ~{\rm MHz}$. As the tip moves closer to the resonator, the scatterer size increases. As a result, either a crossing or an avoided crossing in frequencies of the two split modes may take place depending on the position of the tip. Both cases manifest themselves as an increase in the splitting as seen in Fig. \ref{fig5}(a) \cite{Jiangang2}. The results indicate that the active microcavity and heterodyne measurement scheme can be used for nanoparticle detection.

In order to have an idea about the detection limits of mode splitting in active microcavities in the lasing regime,
we consider a single spherical polystyrene (PS) particle with radius $R$ and refractive index $n=1.59$ deposited on
a resonator of $V=300 ~\mu m^3$ at $f(r)=0.2$. In the limit of large scatterers, the system is dominated by the scatterer induced linewidth broadening $\Gamma_R$. Then the strong coupling condition $2|g|>\Gamma_R$ implies that $(\frac{\lambda_0}{2\pi R})^3>\frac{2(n^2-1)}{3(n^2+2)}$ should be satisfied, where $\lambda_0$ denotes the resonance wavelength. Thus the largest PS particle which induces a detectable splitting is found as $R\sim 406~{\rm nm}$ for $\lambda_0=1550 ~{\rm nm}$. In the case of small scatterers, on the other hand, the system is dominated by $\omega_0/\emph{Q}$ of the lasing mode, i.e., $\Gamma_R\ll\omega_0/\emph{Q}$, implying that $2|g|>\omega_0/\emph{Q}$ should be satisfied. In case of  PS particles with $R=0.5~{\rm nm}$, $\Gamma_R$ and $2|g|$ are calculated as $26 ~{nHz}$ and $13.7 ~{\rm Hz}$, respectively. The value of $2|g|$ is larger than the reported lasing linewidth in WGM resonators \cite{YangTao,narrowlinewidth}. Thus, we estimate that using the lasing mode and heterodyne detection of mode splitting, dielectric particles as small as $1 ~{\rm nm}$ in diameter can be detected which is beyond the reach of passive microcavities.

In conclusion, we demonstrated mode splitting in active microcavities. Contrary to passive microcavities where mode splitting is detected from transmission spectrum obtained by wavelength scan, active cavities allow detection and measurement of mode splitting from a heterodyne beat note formed by mixing the split lasing modes in a photodetector. Thus, our scheme provides a wavelength-scan-free, compact, and inexpensive tool with high resolution for detecting intrinsic and/or intentionally introduced external scatterers in the mode volume of the microcavity. We believe this self-referencing and self-heterodyning scheme is a sensitive platform to investigate light-matter interactions. Moreover, the demonstrated manipulation of the split modes using an external probe may be used to realize a dual-line laser with tunable split frequencies.

The authors gratefully acknowledge the support from CMI at Washington University in St. Louis and NSF under Grant No. 0954941. This work was performed in part at the NRF/NNIN (NSF award No. ECS-0335765) of Washington University in St. Louis.

\end{document}